\documentclass[a4paper]{jpconf}
\usepackage{graphicx}
\usepackage{amssymb}
\usepackage{amsmath}
\usepackage{accents}
\begin{document}
\title{Status of MICROSCOPE, a mission to test the Equivalence Principle in space}

\author{Joel Berg\'e$^1$, Pierre Touboul$^2$, Manuel Rodrigues$^1$, for the MICROSCOPE team}

\address{$^1$ ONERA, the French Aerospace Lab, 29 avenue de la Division Leclerc, 92320 Ch\^atillon, France}
\address{$^2$ ONERA, the French Aerospace Lab, chemin de la Huni\`ere, 91123 Palaiseau, France}

\ead{joel.berge@onera.fr}

\begin{abstract}
MICROSCOPE is a French Space Agency mission that aims to test the Weak Equivalence Principle in space down to an accuracy of $10^{-15}$. This is two orders of magnitude better than the current constraints, which will allow us to test General Relativity as well as theories beyond General Relativity which predict a possible Weak Equivalence Principle violation below $10^{-13}$. In this communication, we describe the MICROSCOPE mission, its measurement principle and instrument, and we give an update on its status. After a successful instrument's commissioning, MICROSCOPE is on track for on-schedule launch, expected in 2016.
\end{abstract}

\section{Introduction}

The 20th century gave birth to the Standard Model of Particle Physics (SM) and to General Relativity (GR). The former, based on quantum physics, describes the world at small scales and high energy, whereas the latter describes gravitation at large scales. Although superbly tested and confirmed in most of their respective regimes, SM and GR are still incompatible, and attempts to unify them have so far failed. Moreover, the discovery of the acceleration of the Universe's expansion put an extra pressure on current physics: whether it stems from dark energy or from a fifth force, or it is due to the failure of GR on cosmological scales, is still unknown. Thus, premiere tests of gravity are underway or planned, using such diverse probes as cosmological probes (e.g. weak lensing, baryon acoustic oscillations with surveys like Euclid\footnote{http://sci.esa.int/euclid/
}, DES\footnote{http://www.darkenergysurvey.org/} or LSST\footnote{http://www.lsst.org}), gravitational waves detection and characterization (e.g. VIRGO\footnote{http://www.ego-gw.it}, LIGO\footnote{http://www.ligo.caltech.edu/} and eLISA\footnote{https://www.elisascience.org/}), or search for Equivalence Principle violations. Looking for the latter is motivated by predictions from various theories aiming to unify GR and quantum physics, or to modify gravity in order to account for the accelerated expansion of the Universe. For instance, string-theory-inspired theories predict a violation of the Equivalence Principle at a level between $10^{-18}$ and $10^{-13}$ \cite{damour12, overduin12}.

MICROSCOPE (Micro-Satellite \`a tra\^in\'ee Compens\'ee pour l'Observation du Principe d'Equivalence \cite{touboul09, touboul12}) is a drag-free microsatellite which aims to test the WEP down to the $10^{-15}$ level, expected for launch in 2016. In this paper, we first motivate the science case for MICROSCOPE in Sect. \ref{sect_wep}; we then give a short overview of MICROSCOPE in Sect. \ref{sect_mic} before stating on its current status in Sect. \ref{sect_status}.

\section{Weak Equivalence Principle} \label{sect_wep}

\subsection{Weak Equivalence Principle and General Relativity}

The Weak Equivalence Principle (WEP) states that two bodies in the same gravitational field experience the same acceleration, independently of their mass and composition. In other words, it states the universality of free fall. Together with the local position invariance (laws of physics do not depend on the position) and the local Lorentz invariance (laws of physics do not depend on the speed of the observer in an inertial frame), this principle constitutes the Einstein Equivalence Principle (EEP). The latter is at the basis of Einstein's General Relativity. Any violation of the WEP (and of the EEP thereof), would indicate that GR is not the ultimate theory of gravity.

\subsection{Tests of the Weak Equivalence Principle}

\begin{figure}[h]
\includegraphics[width=21pc]{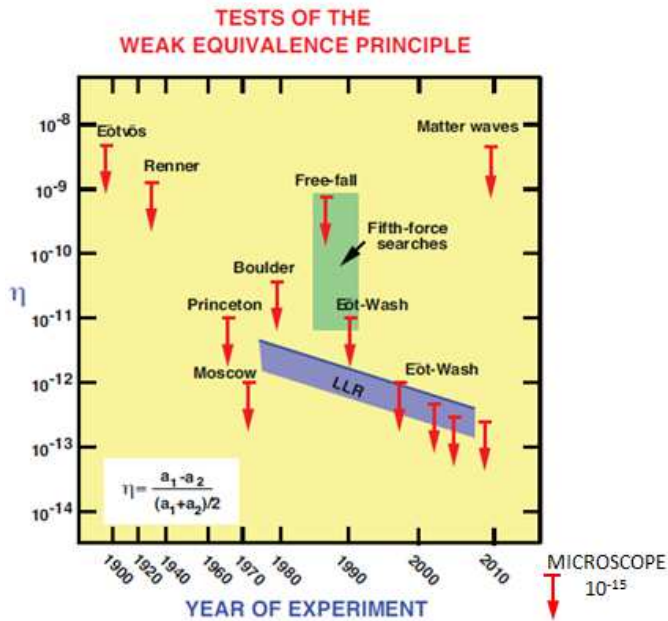}
\begin{minipage}[b]{11pc}\caption{\label{fig_tstep}Tests of WEP throughout the 20th century. The arrow on the lower right corner shows the expectation for MICROSCOPE.  Figure adapted from \cite{will06}.}
\end{minipage}
\end{figure}

The WEP has been tested throughout the 20th century with an increasing precision. Deviations from WEP are usually described by the E\"otv\"os parameter 
\begin{equation}
\eta = 2 \frac{(m_g/m_i)_A - (m_g/m_i)_B}{(m_g/m_i)_A + (m_g/m_i)_B}
\end{equation}
where ``A'' and ``B'' are two bodies experiencing the same gravitational field, $m_g$ is the gravitational mass and $m_i$ the inertial mass. If the WEP holds, then $m_g=m_i$ for all bodies, and $\eta=0$. Figure \ref{fig_tstep} shows how the upper limit on $\eta$ has decreased during the 20th century, as more accurate experiments have been put forth. The first measurement was made by E\"otv\"os with a torsion pendulum, and allowed him to constrain the WEP at a level of $10^{-8}$. Most recently, the monitoring of the Earth and Moon system with the Lunar Laser Range (LLR -- e.g. \cite{murphy12, williams12}), and the measurement by the E\"ot-Wash group with a torsion pendulum, reached the best limits $\eta \leqslant 10^{-13}$ \cite{wagner12}. On-ground measurement are however reaching their limits in terms of signal-to-noise ratio, making a better measurement more difficult. As a consequence, efforts are underway to perform WEP tests with atomic interferometers (e.g. \cite{bonnin13}).
Another solution to increase the precision on the WEP is to test it in space: this is the goal of the MICROSCOPE mission.
Theories currently developed to explain the accelerated expansion of the Universe, or to unify GR and quantum physics, predict that WEP is violated at a level $10^{-18} \leqslant \eta \leqslant 10^{-13}$ \cite{damour12, overduin12}. MICROSCOPE will be able to measure $\eta$ at the level $10^{-15}$: it will thus allow us to probe a significant part of the $\eta$-space under consideration by those new theories, and to start discriminating against theories.

\section{MICROSCOPE} \label{sect_mic}

MICROSCOPE will test the WEP by comparing the acceleration experienced by two free-falling test masses in the Earth's gravity field.
To this aim, it embarks two ultrasensitive electrostatic differential accelerometers. Each accelerometer consists of two coaxial cylindrical test masses whose motion is electrostatically constrained. In one (reference -- `REF') accelerometer, the test masses are made of the same material to demonstrate the experiment's accuracy; they are made of different materials in the second (`EP') accelerometer, which is used to test the WEP. The difference of electric potentials applied to keep the masses in equilibrium is a measure of the difference in the proof masses motion; hence, a non-zero difference of applied potentials is a measure of a WEP violation.

\begin{figure}[h]
\includegraphics[width=21pc]{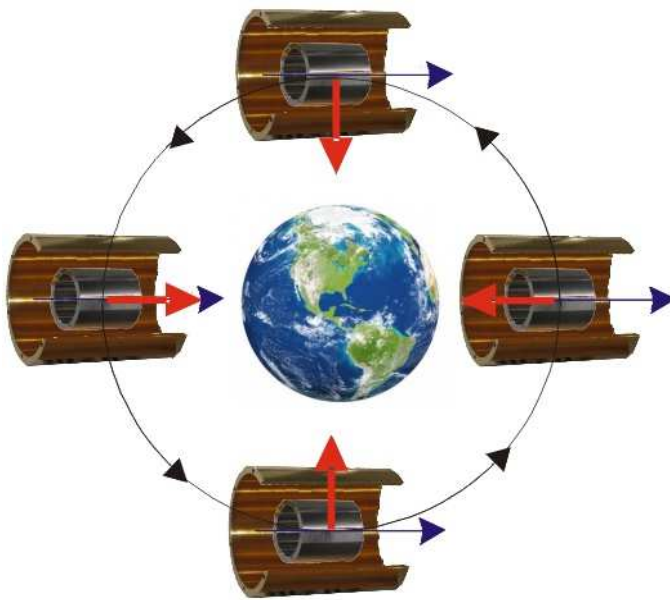}
\begin{minipage}[b]{11pc}\caption{\label{fig_princ}MICROSCOPE's measurement principle. A WEP violation is detected if the two cylindrical test-masses experience different accelerations (red arrows) as the satellite orbits the Earth; the difference in those accelerations is measured by the difference in the voltages applied to the test-masses to keep them in equilibrium. Black arrows show the sensitive axis along which a WEP violation is looked for.}
\end{minipage}
\end{figure}

Figure \ref{fig_princ} shows the measurement principle for the EP accelerometer: its two test masses, since they have the same center-of-mass, experience the same gravitational field (red arrows). If the WEP is violated and if, for example, the internal test mass falls faster than the external test mass, then the difference in accelerations along the EP test axis (horizontal black arrows, along which the test is performed) will be modulated by the instrument's motion around the Earth. We then expect to detect a sine wave corresponding to the modulation of the difference of the voltages applied in the two test-mass electrostatic configuration to keep them centered. Depending on the spacecraft's spin (either null for an inertial session as depicted by the figure, or non-null for a session where the satellite is forced to spin around the axis perpendicular to the orbital plane), the WEP violation signal will have a typical, expected frequency.

The difference of acceleration deduced by the measured difference in applied voltages can be written as:
\begin{multline}
\overrightarrow{\Gamma}_{\rm meas, d} = [\mathcal{M}_c] \left(\eta \overrightarrow{g} + \left( [\mathcal{T}] - [\mathcal{I}n] \right) \overrightarrow{\Delta} - 2 \, [\Omega] \accentset{\bullet}{\overrightarrow{\Delta}} -  \accentset{\bullet\bullet}{\overrightarrow{\Delta}} \right) \\
	+ \overrightarrow{K_{0,d}} + [\mathcal{M}_d] \overrightarrow{\Gamma}_{\rm App,c}
	+ \overrightarrow{\Gamma}_{\rm measquad,d} + \overrightarrow{\Gamma}_{\rm n,d}
	+  [\mathcal{C}_d] \accentset{\bullet}{\overrightarrow{\Omega}}
\end{multline}
where $\overrightarrow{g}$ is the Earth gravity field at the spacecraft's center-of-mass; $\overrightarrow{\Delta} = \overrightarrow{O_iO_j}$ is the distance between the center of the differential accelerometer's test masses; $\overrightarrow{K}_{0,d}$ is the accelerometer's bias; $[\mathcal{T}]$ is the Earth gravity gradient tensor; $[\mathcal{I}n]$ is the instrument's moment of inertia tensor, $[\Omega] \accentset{\bullet}{\overrightarrow{\Delta}}$ is the Coriolis acceleration, with $[\Omega]$ the spacecraft's angular velocity matrix; $\overrightarrow{\Gamma}_{\rm App,c}$ is the common mode acceleration and includes non-gravitational external accelerations; $\overrightarrow{\Gamma}_{\rm measquad,d}$ quantifies the second-order term measured differential acceleration; $\overrightarrow{\Gamma}_{\rm n,d}$ is the instrument's noise; $[\mathcal{C}_d]$ is the differential angular to linear coupling matrix; $[\mathcal{M}_c]$ is the common mode sensitivity matrix and $[\mathcal{M}_d]$ is the differential mode sensitivity matrix. Those last two matrices contain combinations of the accelerometers' scale factor, coupling between axes, and axes alignments with respect to the satellite reference frame, for the common mode accelerations (the half-sum of the accelerations measured by the two accelerometers, which quantifies non-gravitational accelerations) and for the differential mode accelerations (the half-difference of the accelerations measured by the two accelerometers), respectively. They can be finely estimated by in-flight calibrations.
All nuisance parameters are either corrected for through careful modeling (e.g. gravity gradient tensor) or calibrated in flight (e.g. bias \cite{hardy13a, hardy13b}), or minimized by design of the instrument and of the satellite (e.g. inertial tensor, instrument's noise).

MICROSCOPE's instrument (T-SAGE -- Twin Space Accelerometer for Gravitation Experiment) and its performance have been described elsewhere (e.g. \cite{touboul09, touboul12}). As mentioned above, the instrument's mechanical core consists of two differential accelerometers (Sensor Units -- SU), whose test masses are co-axial cylinders kept in equilibrium with electrostatic actuation. The test masses' materials were chosen carefully so as to maximize the scientific return of the experiment and to optimize their industrial machining: the EP test masses are made of alloys of Platinum-Rhodium (PtRh10 -- 90\% Pt, 10\% Rh) and Titanium-Aluminium-Vanadium (TA6V -- 90\% Ti, 6\% Al, 4\% V), while the REF test masses are made of the same PtRh10 alloy.
For each SU, the test masses are controlled electrostatically, through electrodes, without any mechanical contact; only a thin 7 $\mu$m-diameter gold wire, used to fix the masses' electrical potential to the electronics reference voltage, provides a mechanical contact (and associated, accounted for, damping noise) between the test masses and their cage. The test masses' control is performed by an electronic servo-loop. Two Front End Electronics Unit (FEEU) boxes (one per SU) include the capacitive sensing of masses, the reference voltage sources and the analog electronics to generate the electrical voltages applied to the electrodes; an Interface Control Unit (ICU) includes the digital electronics associated with the servo-loop digital control laws, as well as the interfaces to the satellite's data bus. Finally, the same electronics' output is used by the drag-free system of the satellite.

Performance analyses predict a noise Amplitude Spectral Density of $10^{-12}$ m/s$^2$/Hz$^{1/2}$ in the frequency band $10^{-3}-0.03$ Hz, compatible with a test of the WEP at a $10^{-15}$ accuracy (see \cite{touboul09} for a detailed uncertainty analysis).

The spacecraft is derived from the CNES's Myriad series of microsatellites. With a mass of 325 kg and dimensions of 1380x1040x1580 mm$^3$, it has been designed to be as symmetric as possible, with the T-SAGE instrument sitting at its center-of-mass. No moving mechanical parts can contaminate the Equivalence Principle measurement. 
Its propulsion system, based on cold gas, is derived from the GAIA mission and supplied by ESA. It will be adapted and integrated by CNES on the satellite. The Equivalence Principle test will take place on a Sun-synchronous, very low eccentricity, 720 km orbit.

\section{MICROSCOPE status} \label{sect_status}

\subsection{Instrument and satellite}
\begin{figure}[h]
\includegraphics[width=35pc]{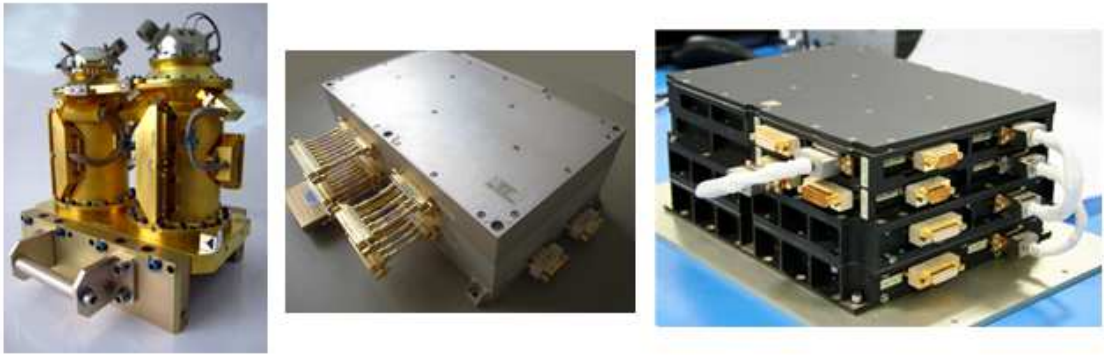}
\caption{\label{fig_instr}MICROSCOPE's Flight Model instrument. From left to right: Sensor Units, FEEU, ICU.}
\end{figure}

As of August 2014, the T-SAGE flight model is assembled. Fig. \ref{fig_instr} shows its Sensor Units, FEEU and ICU.

T-SAGE's Sensor Units have successfully undergone commissioning. In particular, we checked the metrology and verified that the test masses are well aligned. For the EP Sensor Unit, we measured an off-centering of the two test-masses' center of mass of 30.8 $\mu$m, under the required 34.6 $\mu$m. Although we found it slightly bigger than the specification for the REF Sensor Unit (35.7 $\mu$m vs 34.6 $\mu$m), we took it into account to update the error budget and found no appreciable impact on the WEP measurement. All environmental tests (thermal, vibrational) were successfully passed.

Extensive free-fall tests have been performed at the ZARM drop tower in Bremen (see \cite{liorzou14} for their description). In the micro-gravity conditions afforded by those tests, we were able to check that the instrument performs as expected.

T-SAGE's electronics is currently undergoing commissioning. The instrument will be ready for delivery to CNES, where it will be integrated in the satellite, next fall.

\subsection{Science Mission Center}
MICROSCOPE's ground segment operations are centered around three entities: the Command Control Center (CCC) and the Center of Expertise for Compensation of Drag (CECT), hosted by CNES-Toulouse, are in charge of the satellite's operations and drag-free and attitude control respectively, and the Science Mission Center (CMSM), hosted by ONERA-Palaiseau, is in charge of the monitoring and operations of MICROSCOPE's instrument, and of the data processing and analysis. In particular, the CMSM is responsible for (1) ensuring all operational functions to maximize the instrument's operation, (2) day-to-day instrument management and monitoring, (3) weekly mission performance check, (4) proposing modifications of the mission scenario to the Science Working Group, and (5) releasing and archiving the data.

As of August 2014, we are developing the instrument's monitoring and data processing software. Compatibility tests with CNES (that provides preprocessed data) are planned for the fall and winter 2014.

\subsection{Data analysis}
As mentioned above (see also Fig. \ref{fig_princ}), a WEP violation signal will be detected as a sine wave corresponding to the modulation of the difference in accelerations experienced by the instrument's test-masses as the satellite orbits the Earth. The main task of the data analysis is therefore to extract this sine wave from the instrumental noise and to estimate its amplitude. Of foremost importance, we expect a small amount of data to be lost (e.g. because short mechanical crackles will saturate the instrument); as a consequence, the noise will leak from the high frequencies (where its Power Spectral Density increases) to the lower frequencies where the WEP violation signal is measured, therefore hampering the signal detection and estimation. Methods to deal with this problem are currently being implemented: Baghi {\it et al} \cite{baghi15} developed the KARMA (Kalman Auto-Regressive Model Analysis) technique, and Berg\'e {\it et al} \cite{berge15} adapt an inpainting technique (see e.g. \cite{pires09}) to the MICROSCOPE data analysis case.

\section*{Conclusion}
MICROSCOPE will test the Weak Equivalence Principle in space down to an accuracy of $10^{-15}$. This is two orders of magnitude better than the current constraints, and could allow us to rule out new theories that predict a WEP violation around $10^{-14}$, or to complete General Relativity if a WEP violation is detected. Beside this science goal, MICROSCOPE will fulfill a technology objective by showing that the technology is ready for extremely fine satellite attitude control and precise drag-free system. Together with the results of LISA Pathfinder, this will be of interest for future ambitious missions like eLISA.
As the expected launch date is coming, MICROSCOPE's instrument is fully integrated and is close to the end of its commissioning. It will be delivered in the fall 2014 to CNES for integration in the satellite, for a launch on time in 2016. The first scientific results should be available a few months after launch, while the mission is planned to last two years including partial eclipses when the scientific data are not available.

\ack
The authors are grateful to the MICROSCOPE mission staff of CNES in Toulouse, the Observatoire de la C\^ote d'Azur in Grasse and ONERA in Ch\^atillon, for all exchanges about the mission, the satellite, the payload and the science mission center. 
This work was supported primarily by ONERA, where the T-SAGE instrument has been developped under a contract with CNES.
MICROSCOPE is a CNES mission with ESA cooperation.


\section*{References}
\begin{thebibliography}{9}
\bibitem{baghi15} Baghi, Q., M\'etris, G., Berg\'e, J., Christophe, B., Touboul, P., \& Rodrigues, M., Phys. Rev. D. submitted
\bibitem{berge15} Berg\'e, J. et al, in prep
\bibitem{bonnin13} Bonnin, A., Zahzam, N., Bidel, Y., \& Bresson, A.\ 2013, Phys. Rev. A, 88, 043615 
\bibitem{damour12} Damour, T.\ 2012, Classical and Quantum Gravity, 29, 184001
\bibitem{hardy13a} Hardy, {\'E}., Levy, A., M{\'e}tris, G., Rodrigues, M., \& Touboul, P.\ 2013, Space Sci. Rev., 180, 177 
\bibitem{hardy13b} Hardy, {\'E}., Levy, A., Rodrigues, M., Touboul, P., \& M{\'e}tris, G.\ 2013, Advances in Space Research, 52, 1634 
\bibitem{liorzou14} Liorzou, F., Boulanger, D., Rodrigues, M., Touboul, P., \& Selig, H.\ 2014, Advances in Space Research, 54, 1119 
\bibitem{murphy12} Murphy, T.~W., Jr., Adelberger, E.~G., Battat, J.~B.~R., et al.\ 2012, Classical and Quantum Gravity, 29, 184005
\bibitem{overduin12} Overduin, J., Everitt, F., Worden, P., \& Mester, J.\ 2012, Classical and Quantum Gravity, 29, 184012 
\bibitem{pires09} Pires, S., Starck, J.-L., Amara, A., et al.\ 2009, MNRAS, 395, 1265
\bibitem{touboul09} Touboul P. 2009, Space Sci. Rev., 148, 455
\bibitem{touboul12} Touboul P., M{\'e}tris G., Lebat V., Robert A.\ 2012, Classical and Quantum Gravity, 29, 184010
\bibitem{wagner12} Wagner, T.~A., Schlamminger, S., Gundlach, J.~H., \& Adelberger, E.~G.\ 2012, Classical and Quantum Gravity, 29, 184002
\bibitem{will06} Will, C.~M.\ 2006, Living Reviews in Relativity, 9, 3
\bibitem{williams12} Williams, J.~G., Turyshev, S.~G., \& Boggs, D.~H.\ 2012, Classical and Quantum Gravity, 29, 184004
\end{thebibliography}
\end{document}